\documentclass[superscriptaddress,twocolumn]{revtex4}

\usepackage{graphics}
\usepackage{epsfig}
\usepackage{graphicx}
 \usepackage{amsthm}
 \usepackage{color}
\usepackage{amsmath}
\usepackage{xcolor}
\usepackage[latin1]{inputenc}

\begin{document}

\title{Quantitative analysis of the debonding structure of soft adhesives}

\author{Fran\c{c}ois Tanguy}
\affiliation{Laboratoire de Physico--Chimie des Polymeres et Milieux Disperses, UMR7615 ESPCI--CNRS--UPMC,
10 rue Vauquelin, 75005 Paris, France.}

\author{Matteo Nicoli}
\affiliation{Laboratoire de Physico--Chimie des Polymeres et Milieux Disperses, UMR7615 ESPCI--CNRS--UPMC,
10 rue Vauquelin, 75005 Paris, France.}
\affiliation{Laboratoire de Physique de la Mati\`ere Condens\'ee, \'Ecole Polytechnique--CNRS, 91128 Palaiseau, France.}

\author{Anke Lindner}
\affiliation{Laboratoire de Physique et M\'ecanique des Milieux H\'et\'erogenes, UMR7636 CNRS/ESPCI Paristech,
Universit\'e Pierre et Marie Curie, Universit\'e Paris Diderot, 10 rue Vauquelin, 75005 Paris, France.}

\author{Costantino Creton}
\affiliation{Laboratoire de Physico--Chimie des Polymeres et Milieux Disperses, UMR7615 ESPCI--CNRS--UPMC,
10 rue Vauquelin, 75005 Paris, France.}

\begin{abstract}
We experimentally investigate the growth dynamics of cavities nucleating during the first stages of debonding of three different model adhesives. The material properties of these adhesives range from a more liquid-like material to a soft viscoelastic solid and are carefully characterized by small strain oscillatory shear rheology as well as large strain uniaxial extension. The debonding experiments are performed on a probe tack set-up. Using high contrast images of the debonding process and precise image analysis tools we quantify the total projected area of the cavities, the average cavity shape and growth rate and link these observations to the material properties. These measurements are then used to access corrected effective stress and strain curves that can be directly compared to the results from the uniaxial extension.
\end{abstract}
\maketitle

\section{Introduction}

When soft adhesives are detached from rigid surfaces, the incompressibility of the material combined to its extreme deformability leads to
complex deformation patterns involving the formation of air fingers and cavities \cite{Nase:2010,Urahama:1989,Lakrout:1999,Yamaguchi:2007,Zosel:1998}.
The details of these patterns depend markedly on the material properties and
often evolve towards a fibrillar structure of highly stretched polymers which eventually fail by fracture or detach from the surface \cite{Deplace:2009}.
The criteria for the onset of the initial elastic or viscous instabilities have been  known for some time \cite{Crosby:2000,Nase:2008} and several
experimental studies have focused on  fingering instabilities \cite{Nase:2008,Shull:2000,Nase:2011}, on the cavitation criteria \cite{Chiche:2005,Chikina:2000,Poivet:2003,Poivet:2004,Tirumkudulu:2003},
cavity nucleation rate \cite{Peykova:2010,Peykova:2012} or growth rate \cite{Brown:2002}.
However the transition from growth of individual cavities to the collective growth of a population of cavities under the applied stress, leading to elongated
walls between cavities, also called "fibrils" and to eventual detachment, has received much less attention experimentally \cite{Peykova:2010,Peykova:2012}.
Some theoretical papers have been published on collective growth \cite{Yamaguchi2006,Yamaguchi:2006b}.

Up to date, it remains difficult
to relate
the observed patterns to the rheological properties of the soft adhesives, mainly due to the lack of precise experimental characterization of the 3D structures and of the material deformation during the debonding process. Because the
processes are dynamic, powerful 3D scanning techniques, such as confocal microcopy, are too slow and one has to rely on classical 2D imaging limited by  its depth of field.
Proper identification of the cavity borders in an automatic and reliable way it is not a trivial task and requires good quality well contrasted images and
adapted imaging software tools. Yet, this information, albeit statistical in nature, is essential if one wishes to gain more insights on the debonding process and
be able to compare experiments with results from numerical simulations.
It is also a necessary ingredient to understand which rheological properties of the material determine the debonding patterns
and, eventually, the adhesion performance, the important parameter for practical applications.

In this paper we have performed careful experiments yielding high contrast images of the cavities nucleated in the early stages of debonding during a probe tack test. We have developed precise image analysis tools to characterize quantitatively and in a statistically significant way the size, shape and overall projected surface of the cavities. Using model materials with well
known rheological properties spanning from viscoelastic liquids to viscoelastic solids, we will present detailed measurements of the growth dynamics of cavities, including the total projected area, the average cavity shape and their growth rate. These measurements give access to a corrected true stress and strain which can then be quantitatively compared with material properties in shear and uniaxial extension.

\section{Materials and Methods}

\subsection{Sample preparation}

Model acrylate random copolymers containing 98.1\% of butyl acrylate and 1.9\% of acrylic acid and with
varying molecular weights were synthesized by emulsion polymerization by Dow Chemical Company.
Two series (A and B) were produced by changing the conditions  of synthesis in order to obtain different
architectures and, thus, a wide range of viscoelastic properties. Chain Transfer Agent (CTA) was used to control
the weight average  molecular weight, Mw, of the samples for a given series.
For each sample we characterized its Mw and gel content.
The average particle size was found to be around $380$ nm, see Table~\ref{tab_mat}.
It was determined by The Dow Chemical Company that all polymers have a low proportion of short branches.
\begin{table}[t!]
\begin{tabular}{cccccc}
\hline
\hline
 Polymer & CTA  & Mw & PDI & $d_0$ & Gel content \\
  & (\%)  & (kg/mol) & (-) & (nm) & (\%)   \\
\hline
 Bg1110 & - & 1115 & 3.39 & 368 & 30  \\
\hline
 A1570 & -  & 1572 & 2.57 & 400 & -   \\
\hline
 A650 & 0.1  & 651 & 2.18 & 400 & -  \\
\hline
\hline
\end{tabular}
\caption{Properties of the model acrylic polymers. The parameter $d_0$ is the diameter of the particles, see the main text for additional details.}
\label{tab_mat}
\end{table}

Latex solutions have been dried using two different techniques for mechanical  and adhesion tests described below.
Rheology and tensile tests required  thick films  ($\sim$600 $\mu$m), so that latexes were cast in silicone molds and dried
during a week at room temperature  followed by 5 min at 110 $^\circ$C in an oven. For adhesion tests, thin films ($\sim$140 $\mu$m)
coated on glass slides were made. In this case, latexes were cast on glass slides and dried for 24 hours at room temperature followed
by 2 min at 110 $^\circ$C in an oven. In both cases transparent cohesive films were obtained showing a good coalescence of the particles of the latex.

\subsection{Rheology and Probe Tack Test set-up}

The characterization of the viscoelastic properties of the polymers was done on an ARES rheometer (TA) with a standard plate geometry at the Université Catholique de Louvain (UCL).
The frequency range was between $10^{-2}$  and $10^2$ rad s$^{-1}$, while the temperature ranged from 30 to 90 $^\circ$C.
We also performed tensile tests  to obtain the mechanical properties of the sample in uniaxial deformation.
Experiments were carried out in a standard tensile Instron equipment (5565) equipped with a videoextensometer (SVE).
We imposed two different cross-head velocities $v$,  $1.05$  and $0.105$ mm s$^{-1}$, for samples with an initial length $l_0$ of $15$ mm (initial cross section $S_0=2.5$ mm$^2$), resulting in a nominal
initial strain rate $v/l_0$ of  $0.07$ and $0.007$ Hz, respectively.

A home built "probe tack" set up \cite{Josse2004} was employed  to observe the deformation structure of the soft adhesives  and  to measure force and displacement during debonding.
The
apparatus  consists of a cylindrical flat ended probe brought into contact with an adhesive layer. After a  contact time of $10$ s, the probe was pulled away at a constant rate
$V$ of $1$ or $10$ $\mu$m s$^{-1}$. As the thickness of the sample $h_0$ is 140 $\mu$m, the nominal strain rate approximated by $V/h_0$ was
$0.007$ and $0.07$ Hz, respectively. The force $F$ and displacement $d$ were measured during the whole experiment. The probe was made of stainless steel, a material that offers a high surface energy and leads to good adhesion.
Furthermore, in order to obtain a perfectly smooth and reflective surface, the probe was mechanically polished.
In all the experiments  we used a probe with a diameter of $6$ mm.

A microscope was coupled to this experiment in order to observe the debonding structure from the top.
A camera (resolution of $1292\times 964$ pixels) numerical recorded the digitalized images.
Two Zeiss lenses (1.25x and 5x) were used in order to get low or high magnification images, with a field of view of
$7.34\times 5.48$ mm and $1.92\times 1.44$ mm, respectively.
Images and force-displacement data were synchronized with a trigger
to start simultaneously the probe-tack experiment and the image acquisition process.
This trigger also controlled the frequency of the acquisition of the images, setting a frame rate of
$10$ and $20$ fps for a velocity of $1$ and 10 $\mu$m s$^{-1}$, respectively.
\begin{figure}[t!]
\begin{center}
\epsfig{clip=,width=0.5\textwidth,file = 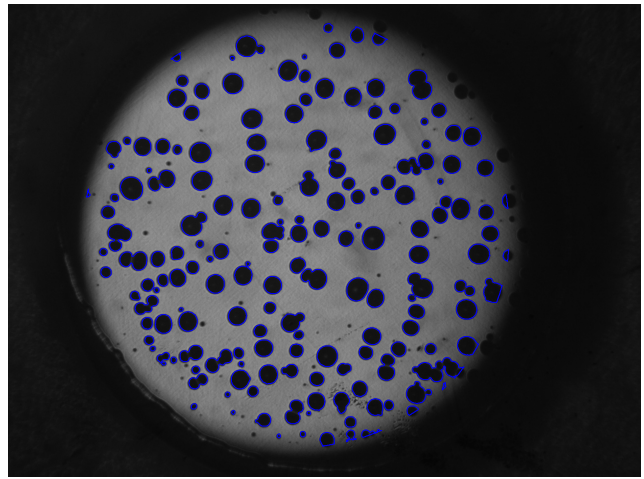}
\caption{Processed top-view frame of the cavitation process. Blue contours  represent the borders of the cavities our
algorithm is able to detect. Only a few small cavities are missed  because they  are  below the noise level.
They will be tracked in  the next  frames when their  area exceeds   $\epsilon_A$ (see the main text).}
\label{fig_image}
\end{center}
\end{figure}

\subsection{Image analysis}

Quantitative information about  the nucleation and the growth of cavities  can be obtained   by
processing the digitalized top-view images acquired in probe-tack experiments.
We developed a simple  method to analyze these images by only resorting to standard routines already
available in many packages for image processing, such as the Image Processing Toolbox\textsuperscript{\texttrademark}
for Matlab\textsuperscript{\textregistered}. An example of the result of this procedure is shown in Figure \ref{fig_image}.

The algorithm detects all cavities with a surface larger than a threshold $\epsilon_A = 50$ pixels. Several geometrical quantities, such as the center of mass, the area, the equivalent diameter and the eccentricity are measured for each cavity. The program also assigns an index to each cavity and by comparing the center of mass of cavities between two subsequent frames, nucleation and coalescence events can be tracked. For additional details see Appendix \ref{App_image}.
\begin{figure}[t!]
\begin{center}
\epsfig{clip=,width=0.51\textwidth,file = 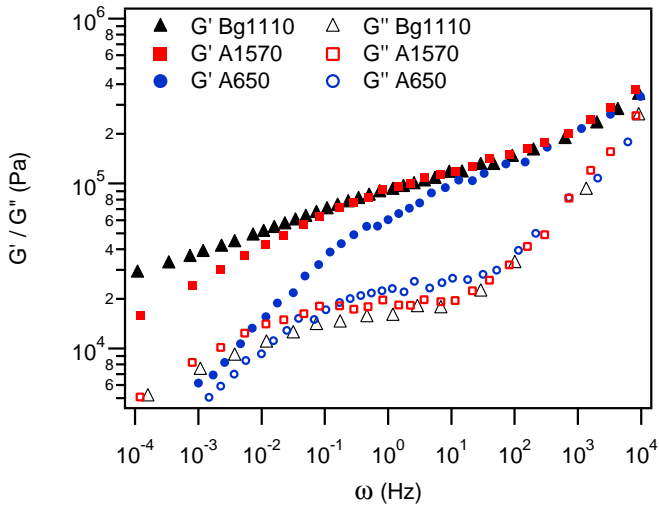}
\caption{Storage ($G'$) and shear ($G''$) modulus as function of  angular frequency ($\omega$) for the three different  materials \cite{Mohite:2013}.}
\label{fig_rheo}
\end{center}
\end{figure}

\section{Results and Discussion}

\subsection{Mechanical Properties}

The three different materials studied differ only in architecture and molecular weight and the molecular interactions with a substrate should thus be the same for all three materials. Figure \ref{fig_rheo} shows master curves at $20^\circ$C of $G'$ and $G''$  as a function of angular frequency $\omega$ \cite{Mohite:2013}. The curves were obtained by applying time-temperature superposition and it can be seen that the viscoelastic properties of the three materials are identical at frequencies $f$ larger than 10 Hz. However, at low frequencies the rheology of A650 differs from the behavior found for A1570 and Bg1110. The elastic modulus of A650 decreases strongly at low frequency, leading to a material with a pronounced viscoelastic character. A1570 and Bg1110 on the other hand can be described as soft viscoelastic solids over the whole range of frequencies. No terminal flow was detected  for any material within the range of frequencies investigated.

\begin{figure}[t!]
\begin{center}
\epsfig{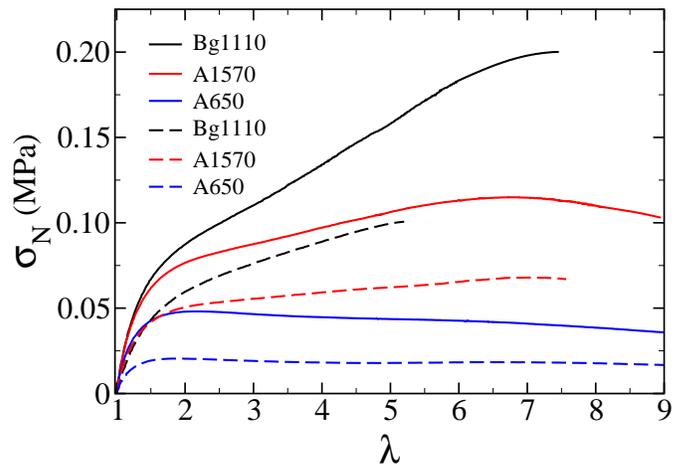}
\caption{Nominal stress versus deformation in tensile test for a deformation rate
$\dot{\lambda}_z = 0.07$ (solid lines) and $\dot{\lambda}_z = 0.007$ (dashed lines).}
\label{fig_tensiles}
\end{center}
\end{figure}

While linear viscoelastic properties characterize time-dependent relaxation processes, strain-dependent behavior is characterized using large strain properties measured at a given strain rate. In uniaxial extension at a fixed crosshead velocity, the materials show pronounced differences,
as shown in Figure \ref{fig_tensiles} by the experimental curves of
nominal stress $\sigma_N=F/S_0$ versus the deformation of the sample $\lambda=l(t)/l_0$. Macroscopic flow is observed for the most viscoelastic material,
i.e.\ the A650 series, while a slight strain hardening behavior characterizes the Bg1110 adhesive.
Although A1570 and Bg1110 have identical linear viscoelastic properties at  frequencies above $0.01$ Hz, the presence of a gel fraction in the Bg1110 series results
in a different large-strain behavior and we observe a markedly higher stress at large strain.

\subsection{Adhesion properties}

Probe tack tests were carried out at two probe velocities for the three materials. For all experiments, the adhesive films have an initial thickness $h_0$ and are pulled by a cylindrical probe of area $A_T$. Experiments were repeated several times and the force $F_T$ and the displacement $d =h(t) - h_0$ as a function of time were measured. The nominal stress is given by
\begin{equation}
\sigma_N=\dfrac{F_T}{A_T},
\end{equation}
while the nominal deformation reads
\begin{equation}
\lambda=\dfrac{h}{h_0},
\end{equation}
and represents the nominal deformation of the whole sample in the vertical direction.
\begin{figure*}[t!]
\begin{center}
\begin{minipage}{0.495\textwidth}
\epsfig{clip=,width=\textwidth,file = 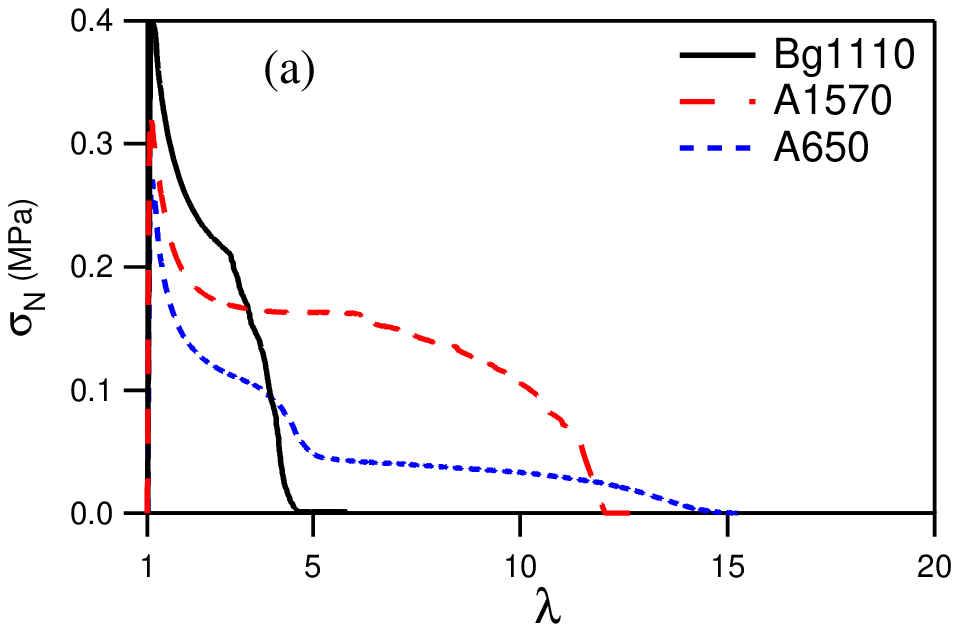}
\end{minipage}
\begin{minipage}{0.495\textwidth}
\epsfig{clip=,width=\textwidth,file = 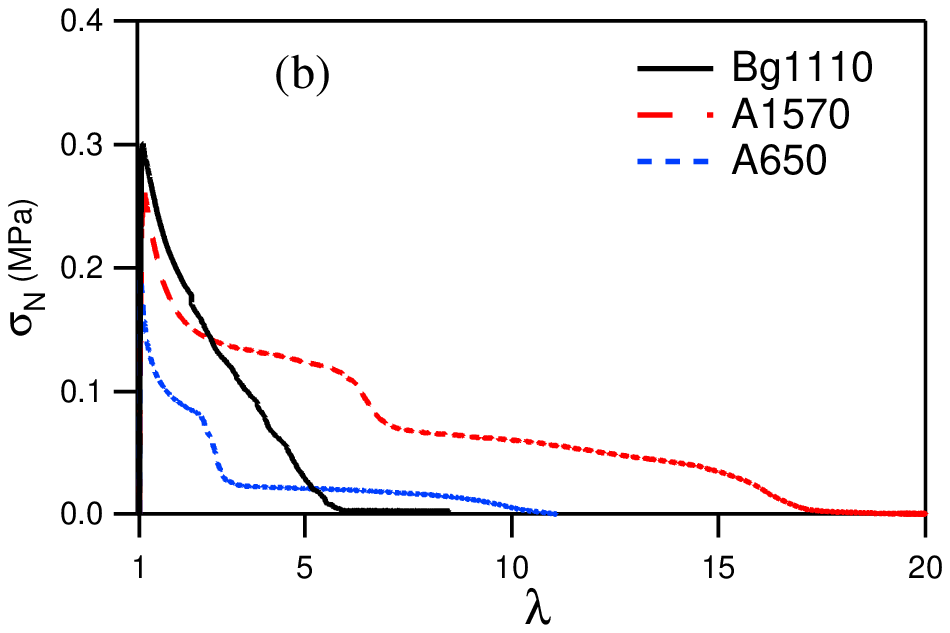}
\end{minipage}
\end{center}
\caption{Nominal stress $\sigma_N$ for the three materials as a function of the deformation $\lambda$ at a pulling velocity of $10$ $\mu$m s$^{-1}$ (a) and
$1$ $\mu$m s$^{-1}$ (b).}
\label{fig_tack}
\end{figure*}

The experiments are filmed at low and high magnification to capture the dynamics of cavity nucleation and growth. During the displacement of the probe the volume between the probe and the glass slide expands. As the adhesive is incompressible and does not slip at the interface, this increase in volume leads to a large increase in tensile stress inside the layer and to the nucleation of cavities at the interface between the probe and the adhesive \cite{Creton:2000} and to their subsequent growth. Note that as the volume of the cavities expands the pressure inside the cavities tends towards zero.

The nominal stress-strain curves $\sigma_N=f(\lambda)$ are shown on Figure \ref{fig_tack}  and are discussed together with the different dynamics of cavity growth. At a debonding
rate of \mbox{10 $\mu$m s$^{-1}$} (Figure \ref{fig_tack}a), three different  shapes  of stress-strain curves are observed for the three materials used. Bg1110,
the most elastic material, shows a sharp stress peak, followed by a fast decrease of $\sigma_N$. This shape is explained by the nucleation of cavities during the increase in $\sigma_N$. These cavities first expand in the bulk of the layer but eventually coalesce at the interface with the substrate. This rapid coalescence leads to the fast decrease in nominal stress observed and results in interfacial  debonding. For A1570, cavities also mainly nucleate during the initial increase of the nominal stress. At higher deformation the nominal stress is found to stabilize at a nearly constant value, characteristic of the growth of cavities in the bulk and the subsequent formation of elongated walls or fibrils. At the end, the fibrils detach from the surface, leading to an adhesive debonding. The experiment with A650 shows a double plateau, characteristic of liquid-like materials. In this case the walls formed between growing cavities are too liquid-like to sustain the pressure difference between the low pressure cavities and the atmospheric pressure and pressure equilibration takes place before final fibril detachment \cite{Poivet:2004}. In this case cohesive failure, i.e. residues on the probe, are observed.

At 1 $\mu$m s$^{-1}$  (Figure \ref{fig_tack}b), the shape of the  stress-strain curve of the  Bg1110 and A650 are qualitatively identical  except for  a decrease of the overall stress during debonding.
For A1570, a transition is observed towards a liquid-like behavior with two plateaus.
\begin{figure}[b!]
\begin{center}
\includegraphics[width = 8cm]{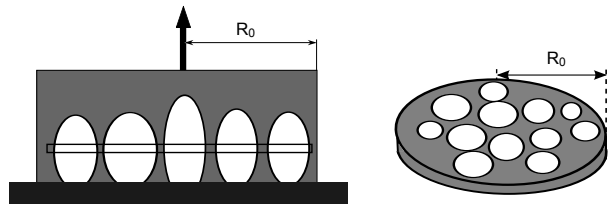}
\caption{Representation of the film under deformation and of the disk considered in the top-view analysis.}
\label{fig_schema_bubbles}
\end{center}
\end{figure}

\subsection{Evolution of the load-bearing area}

Due to the presence of cavities, the force applied on the
disk-shape sample is effectively only applied on a load-bearing cross section that becomes increasingly smaller as $\lambda$ increases.  By analyzing the projected area covered by the cavities and subtracting it from the initial contact area, the average  true stress applied can be calculated (see the following section), instead of the nominal stress studied in previous investigations  \cite{Lakrout:1999,Chiche:2005,Peykova:2010,Zosel:1985}.
\begin{figure*}[t!]
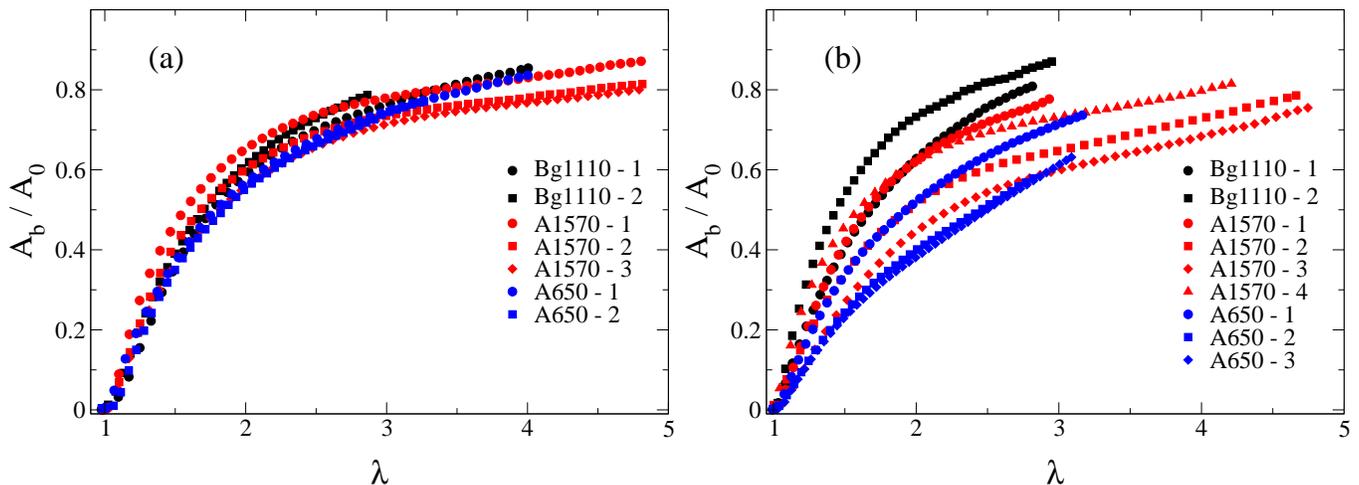

\begin{center}
\begin{minipage}{0.495\textwidth}
\epsfig{clip=,width=\textwidth,file = totalA_v10.eps}
\end{minipage}
\begin{minipage}{0.495\textwidth}
\epsfig{clip=,width=\textwidth,file = totalA_v1.eps}
\end{minipage}
\end{center}
\caption{Evolution of the  ratio $A_b / A_0$ as function of the nominal  deformation $\lambda$. These seven experiments were performed
at a constant  pulling velocity of  10 $\mu$m s$^{-1}$ (a) and 1 $\mu$m s$^{-1}$ (b).}
\label{fig_A}
\end{figure*}

By  means of the image analysis method described in the previous section
we can measure the total contact area  $A_0$ which changes with time. Note that $A_0$ is typically slightly smaller than $A_T$, the area of the probe,  even in the beginning of the experiment. The small difference between these two areas originates from the impossibility, in our setup, to illuminate  the whole surface in contact with the probe. For each frame  we measure the total area covered by the cavities $A_b$ and then deduce the load-bearing cross section of our disk as a function of time, $A_e=A_0 - A_b$. This latter quantity is simply the effective area of the walls between cavities. Note that, as the observation direction is normal to the disk, the maximal diameter of each cavity is observed in the projected image, see  the sketch in Figure \ref{fig_schema_bubbles}.

The precise measurement of the growth dynamics of the cavities can unfortunately not be undertaken for the complete force-displacement curve. Due to loss of contrast and resolution we can only precisely track cavities until $\lambda = 3-5$, i.e.\  the first part the curves shown on Figure~\ref{fig_tack}   and all the following results will be restricted to this deformation range.

The study of the evolution of the projected areas taken by the cavities $A_b$ as a function of time and nominal  deformation $\lambda$ gives interesting insights on the average geometry of the cavities and can be linked to the rheological properties of the material and to the adhesion at the interface with the probe.

 Cavities are observed to nucleate at different locations for each experiment due to the fact that nucleation depends on parameters that are difficult to control, such as dust or  microparticles on the surface. Also the precise moment of nucleation can slightly vary between experiments \cite{Chiche:2005}.
 Interestingly, however, when the probe is pulled at  \mbox{10 $\mu$m s$^{-1}$} the function $A_b/A_0$
(shown on Figure \ref{fig_A}a) is very reproducible for different experiments with the same material and is found to be similar for the three  materials.

On the other hand, when the probe is pulled more slowly (at 1 $\mu$m $s^{-1}$) the picture is  different, see Figure~\ref{fig_A}b.
Contrary to the tests at 10 $\mu$m s$^{-1}$, some scatter is observed for $A_b / A_0$ for each material and $A_b / A_0$ now seems to depend on the material. Bg1110 shows a faster increase in the projected cavity area, then A1570 and A650. This shows that at slow pulling rate, cavities grow with different shapes for the different materials. Cavities growing in the more elastic material cover more surface and are thus less elongated in the tensile direction, whereas cavities in the softer material show a slower increase in projected areas, indicating a more elongated shape. This result is consistent with what was found by
Yamaguchi {\sl et al.} \cite{Yamaguchi:2007} for adhesives with different crosslink densities.

The differences in the measurements  between the two probe velocities are interesting. In fact, they show that at \mbox{10 $\mu$m s$^{-1}$} the shape of the cavity is fully determined
by the high frequency behavior of the materials, which does not vary much between the different materials. On the other hand, at 1 $\mu$m s$^{-1}$, differences in rheological properties do lead to
significantly different kinematics which will eventually lead to very different levels of dissipated energy.
\begin{figure}[b!]
\begin{center}
\epsfig{clip=,width=0.5\textwidth,file = 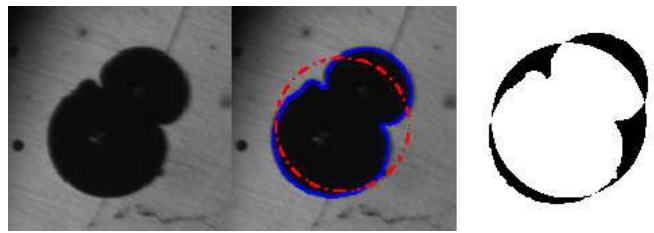}
\caption{From left to right: i) Image of the cavity, ii) Detected perimeter (blue solid line) and equivalent circle (red
dashed line) placed on the center of mass of the cavity, iii) Absolute difference between the two areas  (black region).}
\label{fig_shape_diff}
\end{center}
\end{figure}

\begin{figure*}[t!]
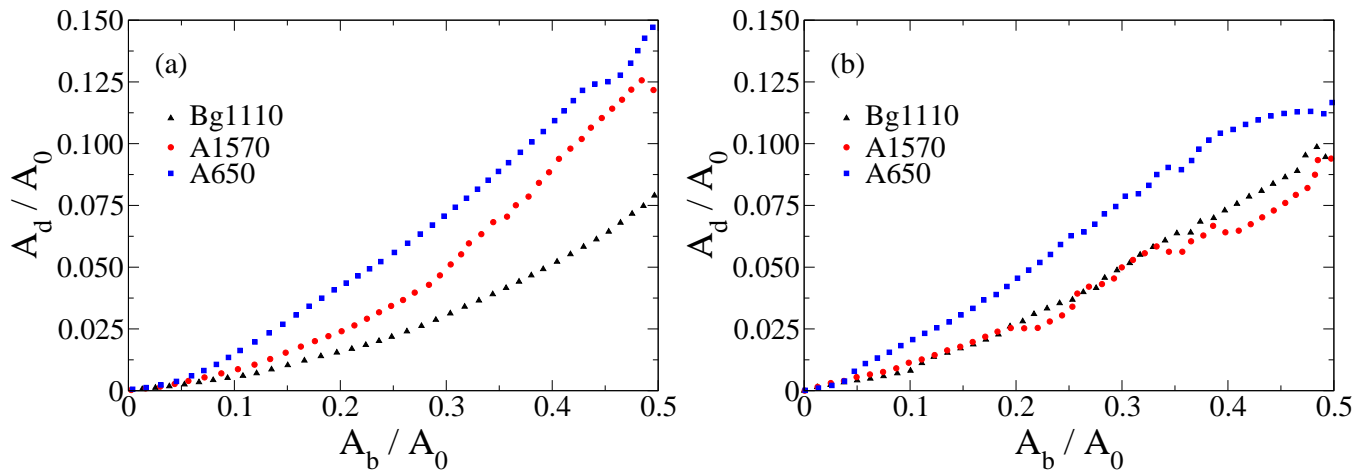

\begin{center}
\begin{minipage}{0.495\textwidth}
\epsfig{clip=,width=\textwidth,file = excess_area_v10.eps}
\end{minipage}
\begin{minipage}{0.495\textwidth}
\epsfig{clip=,width=\textwidth,file = excess_area_v1.eps}
\end{minipage}
\end{center}
\caption{Evolution of the  excess area $A_d / A_0$ as function of the load bearing area $A_b / A_0$
at the  pulling velocity of  10 $\mu$m s$^{-1}$ (a) and 1 $\mu$m s$^{-1}$ (b).}
\label{fig_excess}
\end{figure*}

\subsection{Projected shape of cavities}

During the early stages of the debonding process described in the previous paragraph, the shape of the projected area of individual cavities undergoes a transition
from a circular to a more irregular form. Initially cavities grow in a circular manner. As the cavities start to occupy more volume they begin to feel each other through elastic interactions
and viscoelastic flow. These interactions lead
to a deviation from their initial circular shape and, eventually, to the coalescence of cavities, further modifying the overall shape.
A simple way to quantify this geometrical transition  is to compute the size of the difference between the shape
of the cavity and the circle with the same projected area placed at the center of mass of the cavity, see Figure \ref{fig_shape_diff}.
This absolute difference between areas, $A_d$, provides a  measure of the average change in shape of the cavities, thus quantifying in this way how the material responds to an external deformation. The elasticity of the material acts here like a surface tension and restricts sharp changes in shape \cite{Dollhofer:2004}.

The data is best shown as a function of the relative area occupied by the cavities $(A_b/A_0)$ to avoid effects due
to differential nucleation present for the slower pulling velocity. For all materials and strain rates, the projected area of the
cavities becomes markedly non-spherical as cavities interact or merge with each other.

The evolution of the normalized  $A_d / A_0$  for the two velocities  is shown in Figure \ref{fig_excess}.
At both velocities the more elastic material Bg1110 maintains more circular cavities consistent with its more elastic character. This strongly suggests that  the level of elastic energy stored in the material during deformation has an effect on the curvature of the cavities.

\subsection{Growth rate of individual cavities}

We estimate the growth rate of individual cavities from the evolution of the projected area of each cavity as a function of time shortly after their nucleation.
Images of the whole probe have not enough resolution to provide this information and we thus use high magnification images (5x) of the central part of the sample.

\begin{figure}[t!]
\begin{center}
\epsfig{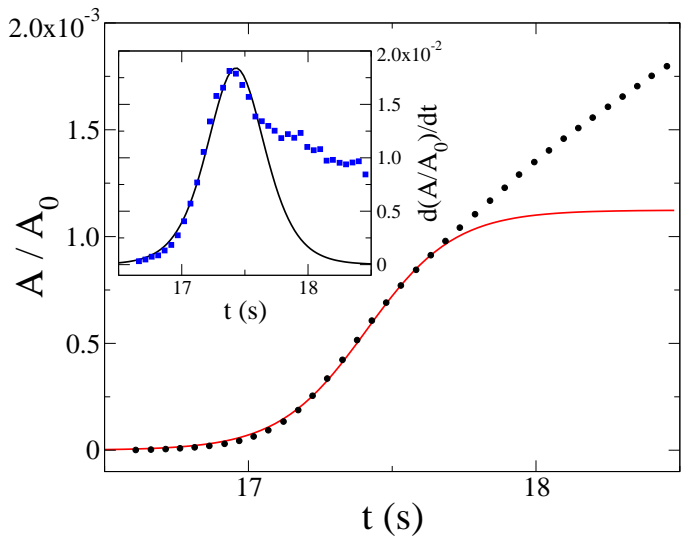}
\caption{Example of the evolution of the area of a growing cavity and its time derivative (inset) as a function of time and example of the fit procedure used to estimate $\tau$. Points are experimental data from digitalized images whereas
solid lines correspond to fits of Eq.\ \eqref{sigmoid} and \eqref{dsigmoid} (inset), respectively.}
\label{fig_fit_procedure}
\end{center}
\end{figure}

The increase in area of a single cavity normalized by the area of the probe is shown as a function of time on figure \ref{fig_fit_procedure}. From this figure it is clear that the growth of cavities does not follow a simple functional form, in agreement with previous observations \cite{Chiche:2005,Peykova:2010,Brown:2002}. Right after nucleation, exponential cavity growth is observed  \cite{Brown:2002}, but quickly after this initial stage they start to interact with the surrounding cavities and their growth slows down and deviates from the exponential behavior. This is easily explained by the fact that cavities relax the accumulated stress in the adhesive layer very quickly after their nucleation, leading to a slow down of the growth.

We aim at capturing the  first stages of cavity growth, as differences between different materials are expected to be important mainly when cavities  grow independently. Even if cavities grow exponentially right after nucleation, the later stages of the growth rate can be approximated by a square root function and a simple exponential fit does not permit a clean estimation of the growth rate $\alpha$. In fact, the time variation of the area $A(t)$ of each cavity reaches a maximum in a very short time and,  subsequently, it decreases.
A sigmoid function $S$ can easily catch this behavior of $A(t)$:
\begin{equation}
S = a \left[1 + e^{-\alpha(t-t_0)} \right]^{-1},
\label{sigmoid}
\end{equation}
where $a$ is the amplitude of $S$ (for $a=1$,  $S\to 1$ when $t\to \infty$), $\alpha$ is the growth rate, and $t_0$
is the moment of maximum growth. These three parameters are estimated from a non-linear
least-squares fit of  the time derivative of $A$ by using the  function
\begin{equation}
\dfrac{dS}{dt} = \dfrac{a\alpha \, e^{-\alpha(t-t_0)}}{\left[1 + e^{-\alpha(t-t_0)} \right]^2}.
\label{dsigmoid}
\end{equation}
A typical result of the fitting procedure is shown in Figure \ref{fig_fit_procedure}.

\begin{figure}[t!]
\begin{center}
\epsfig{clip=,width=0.5\textwidth,file = 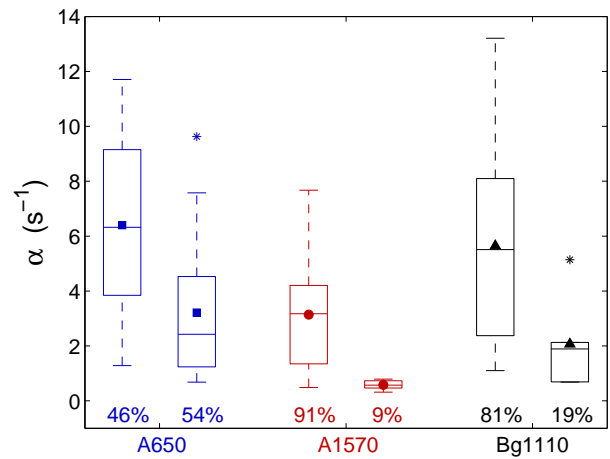}
\caption{Box plot of the growth rate $\alpha$ for the three different materials at the  pulling velocity of  10 $\mu$m s$^{-1}$.
Cavities have been divided into two groups according to their nucleation time,  before (left boxes) and after (right boxes) the force peak.
Percentages  show the proportion of cavities for each group. The total number of cavities were 39 for the A650, 53 for the A1570,
and 32 for the Bg1110. The box plot is characterized by five-numbers summaries, i.e.\ the smallest observation
(the lower horizontal line),
the lower quartile (lower boundary of the box),  the median (the line inside the box), the upper quartile (upper
boundary of the box), and the largest observation (the upper horizontal line).
We have also added the mean of each data set (the symbol inside the box) and outliers are represented by
stars.}
\label{fig_boxplot_growth_rate}
\end{center}
\end{figure}

A box plot of the growth rate $\alpha$ for the three materials is shown in Figure \ref{fig_boxplot_growth_rate}.
We have divided the cavities in two groups, those that have nucleated before  the force peak during
the probe-tack test (left column) and those nucleated after it (right column).

First of all, one can note that for the more elastic materials, Bg1110 and A1570, most of the cavities nucleate before the maximum of the stress peak is reached \cite{Peykova:2012}. For the more liquid-like material however significant nucleation is observed even after the stress peak has been reached. This phenomenon can be explained by the fact that for the low modulus of the A650 material the compliance of the adhesive layer quickly drops below the compliance of the apparatus leading to a sudden transfer of energy from the apparatus to the adhesive layer initiating nucleation of further cavities. This observation might thus be apparatus dependent. The most interesting observation is the difference in growth rate between the different adhesives. The most elastic material, Bg1110, and the most liquid-like material, A650, both show larger growth rates with a large scatter, whereas the growth rate of the A1570 material is found to be smaller and more reproducible. For the Bg1110 the large growth rate of the projected area can be explained by the large amount of elastic energy stored in the elastic layer, leading to strong cavity growth along the interface (a crack propagation mechanism). For A650 the resistance of the material is too small to prevent bulk expansion of cavities, also leading to rapid growth of the projected area. The A1570 material seems to have the optimal material properties and leads to a moderate growth rate. The large scatter in the growth rate of cavities nucleated before the peak, observed for A650 and Bg1100, is most likely due to differential nucleation at different stress levels leading to different growth rates \cite{Chiche:2005,Peykova:2010}. For A1570 the growth rate is dominated by the viscoelasticity of the material leading to smaller differences in the observed growth rates.

\subsection{Effective Normal Stress}

One of the most interesting results that comes from the detailed analysis of the kinematics of deformation is the analysis of the applied force.
The normal component of the force applied to the disk $F_d$ is the sum of  two terms. The first one, $F_m$, arises from the deformation   of  viscoelastic material,
whereas the other contribution, $F_P$, is due to the work done against the atmospheric pressure to increase the volume of the low-pressure cavities
(a suction cup effect).
The force $F_d$ will be simply estimated from the nominal force $F_N$ and
  the ratio  between the area of the  illuminated region and the whole probe area $F_d = F_T A_0 /A_T$.
\begin{figure}[t!]
\begin{center}
\epsfig{clip=,width=0.5\textwidth,file = 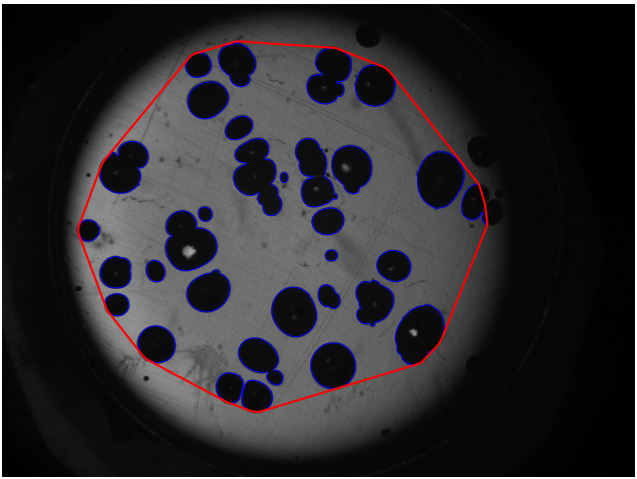}
\caption{Convex envelope of the region occupied by cavities (red solid line) with area  $A_c$.}
\label{fig_convex_hull}
\end{center}
\end{figure}

The fraction of the measured force due to the  work against the atmospheric pressure  depends on the
 spatial distribution of the cavities on the sample. Yamaguchi {\em et al.} used a  simple model
to study the dynamics  of debonding of an axisymmetric PSA simplified to of a one-dimensional problem \cite{Yamaguchi2006,Yamaguchi:2006b}. Their numerical investigations showed
that, after nucleation of cavities, the pressure field rapidly  drops to zero at the position of the two outermost cavities, leading to a
screening effect on other cavities inside the PSA.
This result can be easily extended to our two-dimensional arrangement of cavities by considering the convex envelope of the perimeters
of the cavities. As shown on Figure \ref{fig_convex_hull}, this  area $A_c$ strongly depends on the location of cavities
and can be obtained from the images, so that
\begin{equation}
F_P=A_c (P_{atm}-P_b),
\label{f_bubbles}
\end{equation}
where $P_b$ is the pressure inside the cavities and $P_{atm}$ is the atmospheric pressure.
As $P_b$  is of the order of magnitude of the vapor pressure,  $P_{atm} \gg P_b$ and
equation \eqref{f_bubbles} reduces to \cite{Poivet:2004}
\begin{equation}
F_P \sim A_c \, P_{atm}.
\end{equation}
Although it is obvious that this crude calculation of the pressure field
is not accurate in the nucleation region (before and around the force peak), it gives a good approximation after the force peak
when many cavities are growing simultaneously in size.

From the force $F_m$ we can then calculate the effective tensile component of the  stress applied to the material in the disk
\begin{equation}
\sigma_e = \frac{F_m}{A_e},
\end{equation}
where $F_m = F_d - F_P$.

\subsection{Effective deformation}

The effective stress obtained from the previous calculations is an average value, valid for a slice where the cavities have their maximal diameter
(i.e.\ where the projected area of the walls between cavities is minimum). To plot a true stress versus strain curve, we should also consider the local average deformation
along the tensile direction in the wall for a position in  this slice. This can be obtained from the load-bearing area in a way analogous to the effective stress. The deformation $\lambda=h(t)/h_0$ can be rewritten as $\lambda=R_0^2/R(t)^2=A_0/A_b$ considering volume conservation of a stretched layer without cavity formation. Analogous to the correction of the nominal stress we now use the load bearing area to write the effective deformation as
\begin{figure}[t!]
\begin{center}
\epsfig{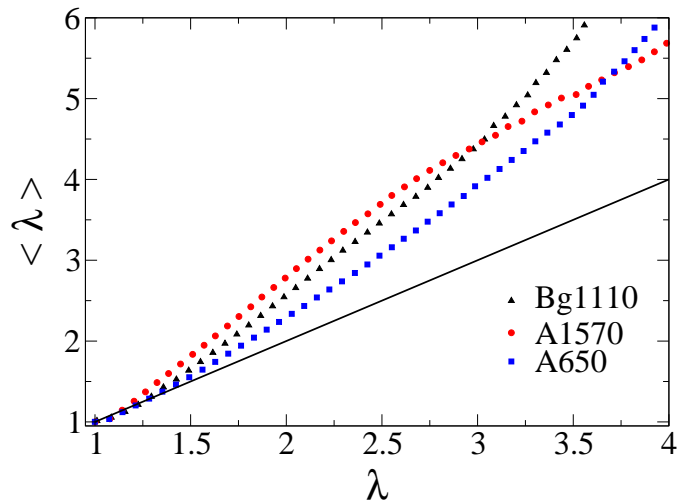}
\caption{Local averaged deformation $\langle \lambda \rangle$ versus global deformation $\lambda$ for the three materials at a pulling velocity of \mbox{$10$ $\mu$m s$^{-1}$}.
The black line is a guide for the eye with slope one.}
\label{fig_local_def}
\end{center}
\end{figure}

\begin{equation}
\langle \lambda \rangle  = \dfrac{A_0}{A_e}.
\end{equation}

In Figure \ref{fig_local_def} this effective average value of $\langle \lambda\rangle$ is plotted as a function of the nominal $\lambda$. The results show that the local deformation always
exceeds the nominal one, suggesting a localization of the deformation in the observation plane analogous to a necking process. The necking
process appears to be unstable (i.e. the slope of $\langle \lambda\rangle$ vs $\lambda$ increases with increasing $\lambda$)  for Bg1110 (crack propagation at the interface due to the stress concentration at the crack tip) and
A650 (no strain hardening and cohesive failure) and stable for the A1570 which has the best PSA properties. This Figure shows well how the
elongational properties of the adhesives should be optimized. If too much elastic energy is stored during elongation, stresses at the edge of the cavities
cannot relax and the cracks coalesce at a relatively low value of $\lambda$. If too little elastic energy is stored, the debonding geometry leads to necking
and cohesive failure. This optimized set of properties is consistent with the PSA design rules proposed by Deplace {\sl et al.} \cite{Deplace:2009} and is also in agreement with the observations made on the growth rates from figure \ref{fig_boxplot_growth_rate}.
\begin{figure*}[t!]
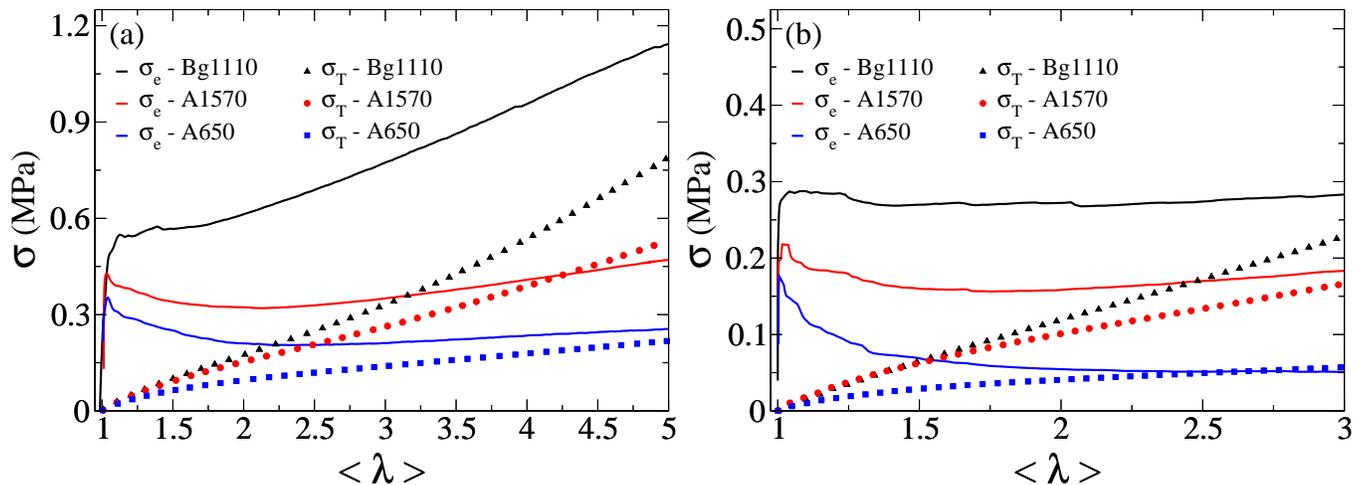

\begin{center}
\begin{minipage}{0.495\textwidth}
\epsfig{clip=,width=\textwidth,file = eff_true_tensile_10micsec.eps}
\end{minipage}
\begin{minipage}{0.495\textwidth}
\epsfig{clip=,width=\textwidth,file = eff_true_tensile_1micsec.eps}
\end{minipage}
\end{center}
\caption{Effective $\sigma_e$ and true tensile  $\sigma_T$ stresses for the three materials at a pulling velocity of $10$ $\mu$m s$^{-1}$ (a) and
$1$ $\mu$m s$^{-1}$ (b).}
\label{fig_eff_true_stress}
\end{figure*}

\subsection{Effective stress vs effective strain curves}

We can now discuss effective stress versus effective strain curves as presented in Figure \ref{fig_eff_true_stress}. The initial peak present in the nominal stress is not
observed anymore for the Bg1110 material and is much less pronounced for the two other materials.
At 1 and  10 $\mu$m s$^{-1}$ the effective stress for the A650 keeps decreasing after the peak and leads, eventually, to cohesive failure.
For the intermediate molecular weight (A1570) the effective stress decreases first and then slightly increases while the most interesting
behavior occurs for the Bg1110 where the effective stress never decreases after the peak force. One would expect the true stress to be much more
directly related to the material properties and it is clear by qualitatively comparing Figure \ref{fig_eff_true_stress} for example with Figure \ref{fig_tensiles},
that the elasticity influences greatly how the effective stress varies with extension. The increase in effective stress for the Bg1110 is clearly related to the
cavities expanding laterally as cracks and this increase in effective stress reflects the presence of a stress concentration at the cavity edge which leads to eventual coalescence of adjacent cavities and debonding. The moderate increase in true stress of the other two materials  is characteristic of the extension of the walls between cavities.

To go even further, we can finally compare the effective stress  $\sigma_e$ as function of $\langle \lambda \rangle$ with the true stress
$\sigma_T= F / A(t)$ (which, due to incompressibility, can be calculated by $\sigma_T= \lambda \sigma_N$)  obtained from the tensile test
(Figure~\ref{fig_tensiles}). Our correction of the stress and strain values from the debonding experiments using the load bearing area is a first attempt to obtain effective stress strain curves that can reasonably be compared to results from material characterization obtained by traction experiments.  The results of this comparison  are shown on Figure \ref{fig_eff_true_stress}a and \ref{fig_eff_true_stress}b. Obviously the two stresses are very different
at values of $\lambda$ close to 1, since the degree of confinement is very high \cite{Crosby:2000,Shull2004}. However, as the deformation of the adhesive layer increases the effective
stress should become closer to the tensile stress in uniaxial extension since the walls between cavities are not confined anymore. This is qualitatively observed in Figure \ref{fig_eff_true_stress}a and \ref{fig_eff_true_stress}b
but one should keep in mind that the stress remains highly heterogeneous in the foam structure and is far from being uniaxial. Note also that for the slow pulling speed the contribution of $F_P$ is more important compared to the faster pulling velocity and small errors made by our approximations might thus be more important for this case. The most striking difference is
between the A1570 and Bg1110 where an apparently small difference in uniaxial constitutive behavior at this strain rate results in a completely different debonding mechanism with a
completely different distribution of local stress.

\section{Conclusions}

We have carried a systematic investigation of the kinematics of deformation of model thin adhesive layers made from acrylic pressure-sensitive-adhesives,
as they are debonded from a flat-ended cylindrical probe at two different probe velocities.

The rheological properties of the three adhesives were characterized in the linear viscoelastic regime and in uniaxial deformation until rupture at two
different strain rates. The three adhesives were chosen to show differences in mechanical behavior at low frequency in small strain and at large
strain due to variable levels of molecular weight and chain branching.

The debonding of the layer from the probe occurred through the nucleation and growth of cavities which then led to an elongated foam structure.
However, the relationship between the applied force and the nominal deformation were markedly different for the three adhesives
representative of behaviors spanning from too liquid-like to too solid-like.

The kinematics of the deformation of the layer was characterized by image analysis as a function of time and the three materials were systematically
compared. The average shape of the cavities nucleating during debonding and the total projected area of the cavities in the plane of the adhesive film
were characterized quantitatively for all three materials at two different velocities. Very few differences in the overall projected area were observed
at $V=10$ $\mu$m s$^{-1}$. However, cavities were more spherical projected area for the more elastic adhesive at
1 $\mu$m s$^{-1}$ while cavities were the most irregularly shaped for the lower molecular weight adhesive. Furthermore an estimate of the local
tensile strain in the plane of observation showed that the local tensile strain systematically exceeds the nominal strain and diverges for the lowest
molecular weight (leading to cohesive debonding) and the most elastic adhesive (leading to interfacial failure by crack propagation) and was only
stable for the intermediate adhesive showing the best PSA properties.

The kinematic information was used to calculate for the first time to our knowledge the effective stress as a function of time in the stage where
cavities grow mostly in the plane of the film and are not yet very elongated in the tensile direction. While this effective stress drops after the peak
force for the two uncrosslinked materials, it keeps increasing after the peak force for the Bg1110. Such a qualitative difference leads to an entirely
different debonding mechanism, with stable fibrils for the two uncrosslinked materials and crack coalescence for the more elastic Bg1110.

These results show that small differences in rheological properties in small and in particular large strain, lead to significant changes in the kinematics
of deformation under the same applied boundary conditions, which then has a great influence on the work on debonding. This coupling between
rheological properties and kinematics is a great challenge for modeling soft materials and we hope that our results will be the base of comparison
with simulations of computational fluid mechanics using realistic material properties.

\section*{Acknowledgement}
{\small
We thank Isabelle Uhl and Ralph Even from the DOW Chemical Company for providing the model adhesives and Lalaso Mohite, Dietmar Auhl and Christian Bailly from the Universit\'e Catholique de Louvain (UCL) for the characterization of the viscoelastic properties of the materials. We acknowledge financial support from the European shared cost project NMP3--SL--2009--228320 MODIFY. \mbox{M.~N.} also acknowledge partial support by
grants  FIS2009--12964--C05--01 and FIS2009--12964--C05--03 (MICINN, Spain), and FIS2012--38866--C05--01 (MEC, Spain).
}

\appendix
\section{Appendix: Image processing}
\label{App_image}

We developed an  image analysis tool based on simple thresholding conversion.
The algorithm starts with a  calibration routine before the nucleation of cavities. In this first step, through a trial and
error procedure, we estimate the critical level $\tau_0$ (with \mbox{$0<\tau_0<1$}) for the conversion from greyscale to binary image.
 Besides, we determine the  region of the image within which  we run our detection routine for the cavities.
This region is established at the beginning of the image recognition procedure and does not evolve with time.
The algorithm assigns  to each cavity an index and manages dynamically the events of nucleation and coalescence [for
details see point 4.].
Obviously, an empty list is  created at the start of the procedure.
Then, the algorithm  repeats the following steps for each frame:
\begin{enumerate}
\item The image is filtered with a low pass filter
in order to reduce its noise content, we typically use a simple averaging over windows of size $3\times 3$ pixels.
\item The format of the image is converted from  greyscale to binary according to $\tau_0$, that is,
all the pixels with luminance smaller  than $\tau_0$ are mapped  to $1$ (white) while the others to $0$ (black).
\item All the  connected regions with area smaller that an threshold $\epsilon_A$ are removed. This step is
easily implemented by morphologically opening   the binary image.
\item The boundaries between black and white regions
are traced and labeled with an index. The children of each parent object are discarded in order to avoid the wrong detection of
small cavities  inside a large encompassing cavity. These white spots are created by the unscattered light that passes  through the cavity and is
reflected back from the steel substrate. Although  their position and their extension  is related to the contact region of the cavity
with the steel substrate,   these quantities
are very sensitive to many  irreproducible factors, such as the intensity of the light, the magnification factor and the sample alignment.
For this reason  these white spots are not taken into account in the analysis of the images from  whole probe experiments.
\item For each cavity several geometrical quantities are measured, e.g.\ its center of mass, area, equivalent diameter, and eccentricity.
\item  By comparing the center of mass of cavities in the current and the previous frame, the index of each cavity is changed according
to the list of indexes of the previous frame. In this step the processes of nucleation and coalescence of cavities are handled. For each
new nucleated cavity a new entry in the list is created with a new index $n_T+1$, where $n_T$ is the largest index of the list.
However, when the coalescence of two or more cavities occurs,
 the new data of the coalesced cavity are assigned to the lowest index in the list while the entries  of the other cavities are deleted.
In this manner we are able to track  the evolution of each
cavity and record all the coalescence events.
\end{enumerate}

\bibliographystyle{unsrt}
\bibliography{tanguy_paper} 
\end{document}